\documentclass[twocolumn]{aastex62}
\usepackage{graphicx}

\received{2019 December 17}
\revised{2020 February 10}
\accepted{2020 February 17}

\begin{document}

%\preprint{}

\title{Orbital Evidences for Dark-matter-free Milky Way Dwarf Spheroidal Galaxies}

\correspondingauthor{Francois Hammer}
\email{francois.hammer@obspm.fr}

\author{Francois Hammer}
\affiliation{GEPI, Observatoire de Paris, Universit\'e PSL, CNRS, Place Jules Janssen 92195, Meudon, France}
\author{Yanbin Yang}
\affiliation{GEPI, Observatoire de Paris, Universit\'e PSL, CNRS, Place Jules Janssen 92195, Meudon, France}
\author{Frederic Arenou}
\affiliation{GEPI, Observatoire de Paris, Universit\'e PSL, CNRS, Place Jules Janssen 92195, Meudon, France}
\author{Jianling Wang}
\affiliation{NAOC, Chinese Academy of Sciences, A20 Datun Road, 100012 Beijing, PR China.}
%\thanks{}
\author{Hefan Li}
\affiliation{School of Physical Sciences, University of Chinese Academy of Sciences, Beijing 100049, P. R. China}
\affiliation{GEPI, Observatoire de Paris, Universit\'e PSL, CNRS, Place Jules Janssen 92195, Meudon, France}
\author{Piercarlo Bonifacio}
\affiliation{GEPI, Observatoire de Paris, Universit\'e PSL, CNRS, Place Jules Janssen 92195, Meudon, France}
\author{Carine Babusiaux}
%\affiliation{Universit\'e de Grenoble-Alpes, CNRS, IPAG, F-38000 Grenoble, France }
\affiliation{GEPI, Observatoire de Paris, Universit\'e PSL, CNRS, Place Jules Janssen 92195, Meudon, France}


%\date{\today}

\begin{abstract}
The nature of Milky Way dwarf spheroidals (MW dSphs) has been questioned, in particular whether they are dominated by dark matter (DM). Here we investigate an alternative scenario, for which tidal shocks are exerted by the MW to DM-free dSphs after a first infall of their gas-rich progenitors, and for which theoretical calculations have been verified by pure N-body simulations. 
Whether or not the dSphs are on their first infall cannot be resolved on the sole basis of their star formation history. In fact, gas removal may cause complex gravitational instabilities, and near-pericenter passages can give rise to  tidal disruptive processes.
Advanced precision with the Gaia satellite in determining both their past orbital motions and the MW velocity curve is however providing crucial results. 

First, tidal shocks explain why DM-free dSphs are found preferentially near their pericenter where they are in a destructive process, while their chance to be long-lived satellites is associated to a very low probability  P$\sim$ 2 $\times$$10^{-7}$, which is at odds with  the current DM-dominated dSph scenario. Second, most dSph binding energies are consistent with a first infall. Third, the MW tidal shocks that predict the observed dSph velocity dispersions are themselves predicted in amplitude by the most accurate MW velocity curve. Fourth, tidal shocks accurately predict the forces or accelerations exerted at half-light radius of dSphs, including the MW and the Magellanic System gravitational attractions.  

The above is suggestive of dSphs that are DM-free and tidally shocked near their pericenters, which may provoke a significant quake in our understanding of near-field cosmology.  
 %, except perhaps if MW dSphs were not representative of dwarfs in the cosmological context.
\end{abstract}

\keywords{Galaxy: general --  galaxies: dwarf -- (cosmology:) dark matter}

%%%%%%%%%%%%%%%%%%%%%%%%%%%%%%%%%%%%%%%%%%%%%%%%%%

%%%%%%%%%%%%%%%%% BODY OF PAPER %%%%%%%%%%%%%%%%%%
\section{Introduction}
\label{intro}
Dwarf spheroidals (DSphs; classical and ultrafaint dwarfs, UFDs, defined here as having $L_V$ $\le$ 2.5 $10^5 M_{\odot}$) in the Milky Way  (MW) halo are by far the smallest galaxies that can be detected and studied. By construction they are the only objects sampling the lower end of the galaxy mass function, which underlines their role in constraining modern cosmology. Their large velocity dispersions have led to assuming that they contain large amounts of dark matter (DM), whose fraction increases with decreasing luminosity or stellar mass \citep{Strigari2008,Walker2009,Wolf2010}. Their preponderant population of old stars has led to the assumption that they are satellites of the MW having reached the halo since early epochs, justifying further the need for large masses to shield them against the destructive MW tidal forces. However, the predictions from the dSph star formation histories (SFHs) may have been overinterpreted, especially without knowing their progenitor properties, their former gas content, and furthermore the way it has been removed from them (see discussion in \citealt{Hammer2019}, hereafter H19). Also, several dSphs, including the most massive Fornax and Sagittarius, show extended SFHs, and building UFD SFHs is still challenging since it requires a sample of at least 200-300 stars near the main-sequence turnoff \citep{Brown2014}. SFHs and their interpretations are not sufficiently compelling for determining the role of DM in dSphs.  \\

The role and presence of DM in dSphs are and have been discussed because of the following questions, perhaps  ordered by increasing importance:
\begin{itemize}
\item Why are some dSphs in a destruction process after only  a few passages (e.g., Hercules; see \citealt{Kupper2017}) while their DM subhalo should have shielded them from tides?
\item Why does it becomes so difficult to distinguish heavily DM-dominated UFDs from DM-free star clusters? As both stellar systems substantially overlap in the $L_V$ - $r_{half}$ plane, a significant fraction of them are not classified up to date, and this is not only due to the quality of the measurements \citep{Simon2019}. This applies also to Crater, which has been included in the dSph sample of \citet[hereafter F18]{Fritz2018}.
\item Why are dSphs so numerous  near their pericenters \citep{Fritz2018,Simon2018,Simon2019}? An observational bias is unlikely to fully explain this since \citet{Simon2019} found that the Sloan Digital Sky Survey (SDSS) and the Dark Energy Survey (DES) should be complete at $L_V$ $\sim$ 21,500 and 3400 $L_{\odot}$, respectively, out to beyond 300 kpc from the MW center.
\item Why are dSphs so numerous in the Vast Polar Structure (VPOS, \citealt{Pawlowski2014})? LMC is just passing its pericenter \citep{Kallivayalil2013}, and it represents a far much larger mass than all dSphs together. LMC debris are fitting well the Magellanic Stream and both the VPOS and a huge fraction of dSphs in the southern Galactic hemisphere (see Figure 4 of \citealt{Kallivayalil2018}).
\item Why can the dark-to-visible matter ratio can be determined by the sole knowledge of the MW gravitational attraction (together with the dSph stellar mass and $r_{half}$; see \citealt{Hammer2018})?
\end{itemize}

Here we show that the above questions can be better addressed in a Newtonian gravitational frame, for which dSphs are not at equilibrium and are in a destructive process caused by the tidal forces of the MW. In Sect.~\ref{dSphorbits} we show that the dSph locations near their pericenters are not consistent with them being long-lived MW satellites, but rather consistent with a first infall together with a destructive process well after the pericenter passage. In Sect.~\ref{tdshocks} we calculate the exact effect of MW tidal shocks, which is further confirmed by pure N-body simulations. 
%We investigate further the predictions proposed by the two scenarios, tidal shocked DM-free or heavily DM-dominated dSphs. We consider the same sample of 21 dSphs as in \citet{Hammer2019}, which has been built from \citet{Fritz2018}. We find that the MW velocity curve predicts accurately the amplitude of tidal shocks for the nearest dSphs, and vice versa, while it appears at odds for the DM-dominated dSph scenario. In the context of a declining rotation curve leading to a moderate MW mass, we show that most dSphs have very eccentric orbits, consistently with a first passage and with the tidal shock scenario. 
In Sect.~\ref{All} we further consider additional stellar systems cataloged in F18, contributing to the debate on the Crater's nature, and explaining why, together with  Hydrus, Reticulum II and Carina II, they have surprisingly low mass-to-light ratio compared to dSphs in the same stellar mass range. We finally examine the properties of dSphs for which only an upper limit on the velocity dispersion has been found, and we propose a new classification scheme that applies for almost all dSphs, through a simple link between their past orbital history and their intrinsic properties.

\section{Calculating dSph orbits consistent with the MW velocity curve}
\label{dSphorbits}
\subsection{Consequences of the MW velocity curve from Gaia DR2}
\label{RC}
For a given set of dSph proper motions, assuming high-mass for the MW generally provides more circularized orbits. The adoption of high mass MW (e.g, model of \citealt{McMillan2017}) by \citet{Helmi2018} has led to moderately eccentric orbits for the classical dSphs and Bootes I. However, after introducing up to 30 UFDs (compared to the nine classical dSphs) the average eccentricity has increased a lot (F18), even when assuming their high MW mass model (average eccentricity of 0.59 for 39 objects vs. 0.44 for the nine classical dSphs). 

\begin{figure*}
	\includegraphics[width=2.0\columnwidth]{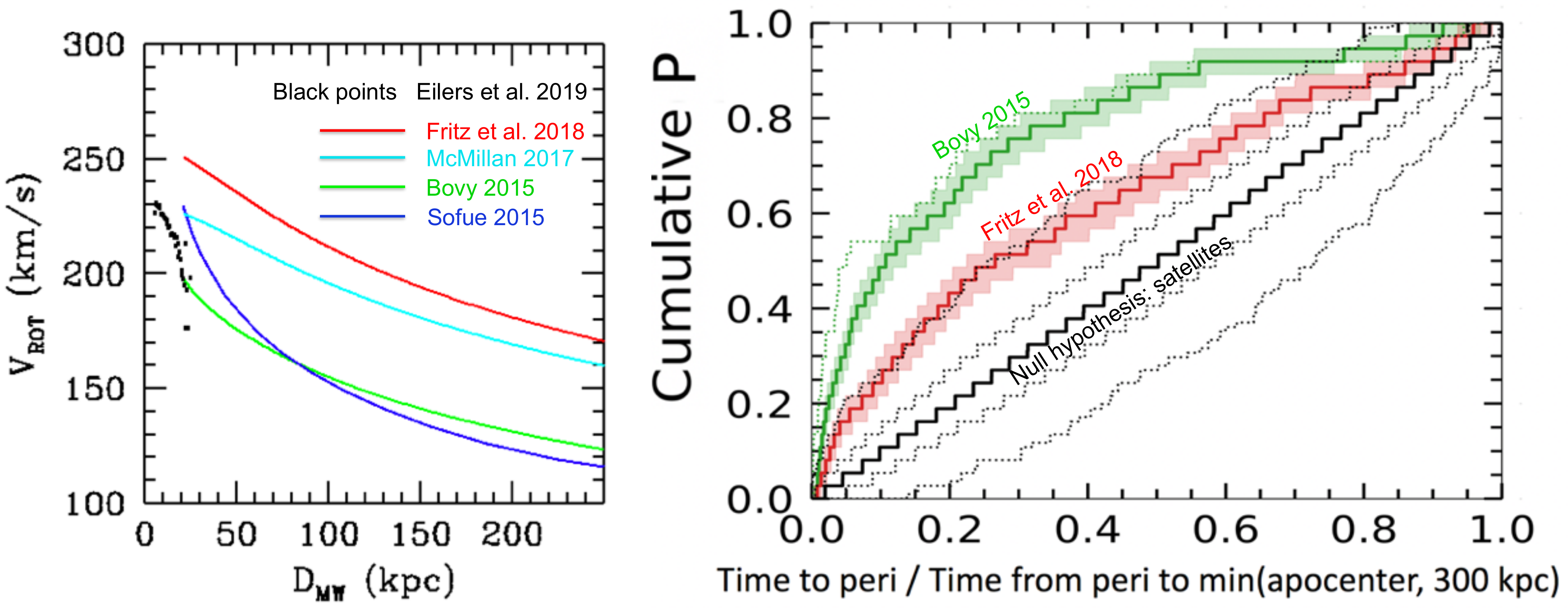}
    \caption{{\it Left:} extended rotation curve of the MW. Comparison of the rotation curve derived from different models to \citet[see small black points]{Eilers2019}.  The green and blue lines shows the \citet{Bovy2015}, and the \citealt{Sofue2015} models, respectively, while the red and cyan lines represent the high MW mass models of (F18, for which they have multiplied by 2 the halo mass of \citealt{Bovy2015}) and of \citet{McMillan2017}, one of the massive MW model used in \citet{Helmi2018}, respectively. For calculating analytically the \citet{Bovy2015} model we have assumed a value of $\rho_{\rm DM}$= 0.0075 instead of 0.008 $M_{\odot} pc^{-3}$ (Bovy, J. 2019, private communication), and twice this value for the F18 model. {\it Right:}  cumulative distribution of time to reach the pericenter divided by the time taken from pericenter to min(apocenter, 300 kpc). The black solid curve shows the median value of the same cumulative distribution based on our simulation that calculates the locations of 1000 satellites per orbit, randomly distributed from pericenter to min(apocenter, 300 kpc). The dotted black curve shows the 1$\sigma$ and 3$\sigma$ limits of that distribution. As in the left panel, the green and red lines represent the \citet{Bovy2015} model and the massive mass model by F18, respectively. Kolmogorov-Smirnov tests provide $D_{max}$ = 0.4667 and 0.225 for the two models, respectively.  }
   \label{fig:1}
\end{figure*}

The circular velocity of the MW has been established with an unprecedented accuracy by \citet{Eilers2019} using 6D phase-space coordinates of 23,000 red giant stars and by \citet{Mroz2019} using 773 Classical Cepheids with precise distances. This leads to an MW mass distribution well described by an axisymmetric and equilibrium\footnote{However, the multiple passages of Sagittarius could have impacted the stellar vertical motions in the MW disk \citep{Laporte2019}, and it is unclear whether the disk rotation could be affected in any manner (see \citealt{Carrillo2019}).} model \citep{Nitschai2019}, which becomes extremely accurate and can be used to test the MW density profile. These results are based on Gaia DR2 \citep{Brown2018} and reveal a gentle but significant decline of the velocity curve ($V_{\rm rot}(D_{\rm MW})$) at MW distances ($D_{\rm MW}$) from 6 to 20-25 kpc (see the left panel of Figure~\ref{fig:1}). The latter range reaches the lower end of the MW distance range for the nearest dSphs. %The MW velocity curve can be well reproduced by a single slope of $\Delta V_{\rm rot}/\Delta D_{\rm MW}$= -1.7 and -1.34 km $s^{-1} kpc^{-1}$ in the 6-20 kpc range, from \citet{Eilers2019} and \citet{Mroz2019}, respectively. \\

The main argument for a high-mass MW originates from comparisons with cosmological models for which MW dSphs are DM-dominated subhaloes. Cosmological models predict that objects such as Leo I or LMC moving fast on eccentric orbits are very rare, implying a very massive MW \citep[mass up to 24 $\times$ $10^{11} M_{\odot}$]{Boylan-Kolchin2011,Boylan-Kolchin2013} to gravitationally bound them. %The same argument also applies to many low mass dSphs that also show very eccentric orbits, and this had led \citet{Fritz2018} to favor a high mass for the MW. %The later arguments could not be advocated against the tidal shock scenario, since it either points out a discrepancy of the DM model at small scales, or leads to a singularity of the MW dwarf system.
However, the left panel of Figure~\ref{fig:1} shows that the low MW mass model of \citet{Bovy2015} is consistent with the \citet[compare the black points with the green line]{Eilers2019} velocity curve from 10 to 20 kpc. The latter excludes high MW mass models, either from F18 or from \citet{McMillan2017}. Another illustration of this can be found in Figure 10 of \citet{Vasiliev2019}, which shows the significant discrepancy between the \citet{McMillan2017} and \citet{Bovy2015} rotation curves at distances larger than 8-10 kpc. We verify that the \citet{Bovy2015} model and that made by \citet{Eilers2019} to reproduce the MW velocity curve do not differ by more than 3 km$s^{\rm -1}$ in rotational velocity. In the following, we will adopt the \citet{Bovy2015} MW mass distribution as our fiducial model, which allows us to use the orbital parameters calculated by F18.\\

\begin{figure*}
	\includegraphics[width=1.8\columnwidth]{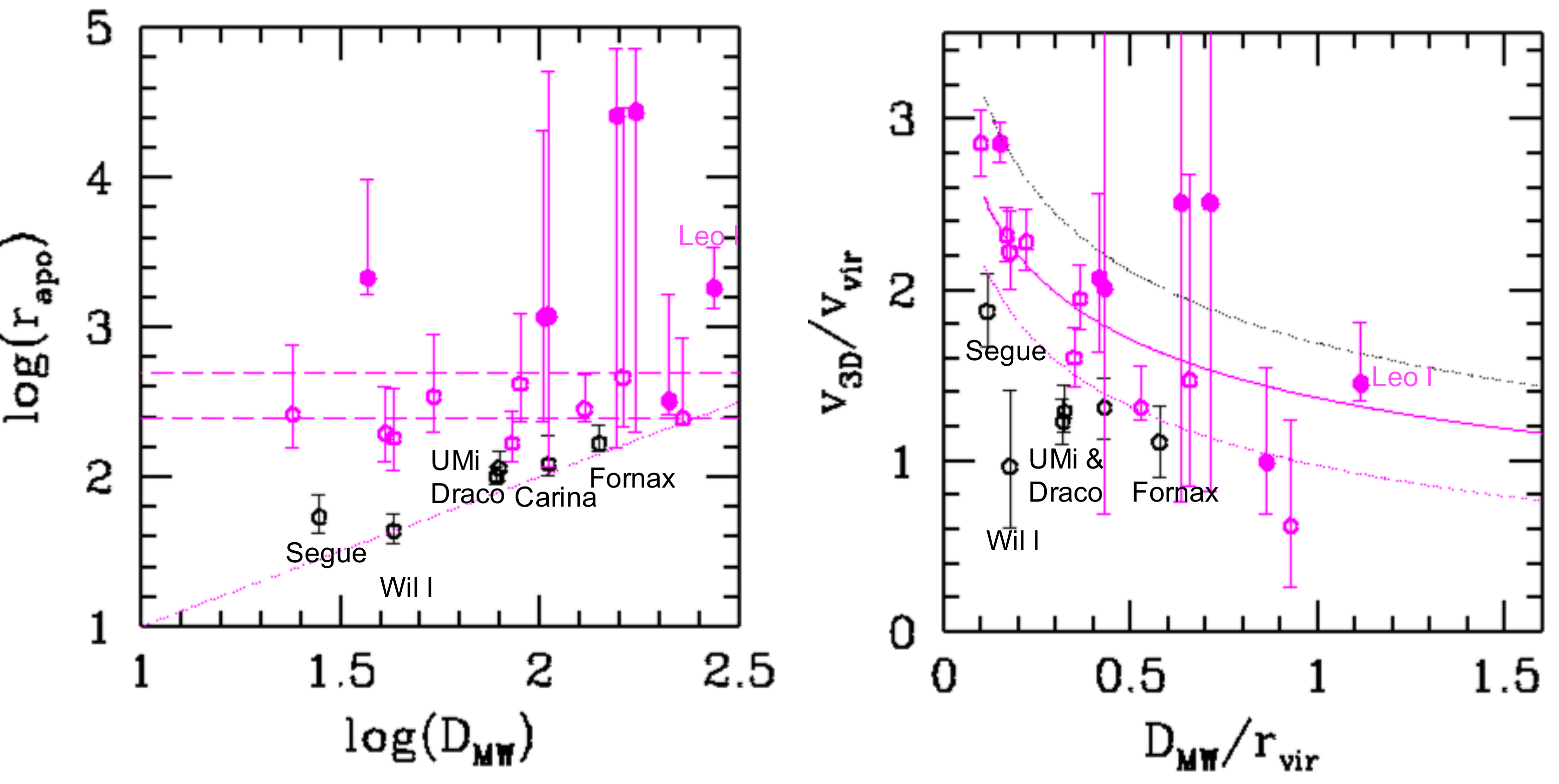}
    \caption{{\it Left:} apocenter distance versus the MW distance ($D_{\rm MW}$). Open black, open magenta, and filled magenta circles represent dSphs fully captured by the MW, dSphs for which the apocenter could be at 1$\sigma$ above the virial radius (245 kpc, enclosing 8 $\times$ $10^{11} M_{\odot}$ for a halo concentration c=15.3 kpc; \citealt{Bovy2015}), and high-orbiting dSphs, for which the minimal value of the apocenter is larger than 245 kpc, respectively. The two horizontal dashed lines mark the 245 and 490 kpc limits, and the dotted line identifies apocenter=$D_{\rm MW}$. {\it Right:} phase-space plots for MW dSphs based on the full determination of 3D velocities by F18. The MW model is from \citet{Bovy2015}, and we have used the color code adopted in the left panel for representing the dSphs. As in \citet{Boylan-Kolchin2013} the solid magenta line represents a curve of constant energy with $v_{\rm 3D}= 1.15 V_{\rm vir}$ and the space between it and the dotted magenta curve delimits the recently accreted subhalos as expected from the {\it Aquarius} simulation (see text). The dotted black line represents the escape velocity, $v_{esc}=\sqrt{-\Phi}$, with the potential $\Phi$ given by \citet[see their Eq.~2.67]{Binney2008}. }
    \label{fig:2}
\end{figure*}

\subsection{Are MW dSphs ordinary satellites?} 
\label{orbits}
\citet[F18]{Fritz2018} and \citet{Simon2018} noticed an excess of dSphs lying near their pericenters, which is at odds with expectations for satellite orbits since most of them should lie near their apocenters, where their velocities are smaller. In other words, if dSphs with eccentric orbits were MW satellites, they should rarely be seen at pericenter, like, e.g., comets in the solar system. \\
 Using the elliptical orbits provided by Table 3 of F18 one could calculate the probability of finding a dSph at a distance from the MW equal to or lower than what is observed, which is also given by the duration to reach (or to leave) the pericenter divided by half the orbital period. In fact, the above calculation would only provide an approximation since actual orbits around an extended mass distribution, such as that of the MW, are more like a rosette than an ellipse (\citealt{Binney2008} and see also illustration in the Appendix B of \citealt{Wang2012}). This has led us to use the publicly available code galpy \citep{Bovy2015} for calculating the dSph orbits from their proper motions (see also F18) and to derive the time spent to reach the pericenter, as well as that from apocenter to pericenter. \\
However, \citet[see also \citealt{Drlica-Wagner2019}]{Simon2019} shows that surveys from SDSS and DES are complete within 300 kpc for dSphs with $L_{\rm V}$ larger than 21,500 and 3400 $L_{\odot}$, respectively. This prompts us to compare the dSph locations on their orbits after limiting them to a maximal extent of min(apocenter, 300 kpc) to that of randomly distributed satellites within the same limited orbits (our null hypothesis). It leads us to exclude Eridanus II and Phoenix (distances in excess of 300 kpc) from the F18 sample considered here, leading to 37 dSphs. To describe further the null hypothesis and its variance, we consider 1000 artificial satellites per dSph orbit, for which their orbit location is just randomly selected from pericenter to min(apocenter, 300 kpc). \\
The right panel of Figure~\ref{fig:1} presents the cumulative probability of having an excess of dSph positions near the pericenter for the \citet{Bovy2015} and F18 models (green and red lines, respectively), when compared to expectations from normally distributed satellites (black solid line). Applying a nonparametric Kolmogorov-Smirnov test leads to a maximal distance of $D_{\rm max}$= 0.486\footnote{ If no truncation at 300 kpc was made, it would have led to $D_{\rm max}$= 0.54; see the green dotted line in the right panel of Figure~\ref{fig:1}} and 0.27 for the two models, respectively. %We assume for the null hypothesis that all dSphs are MW satellites, i.e., their location in their orbital path is at random and shows no preference to be near pericenter (black line in Figure~\ref{fig:1}). 
The associated probability that the observed excess of dSphs near pericenter is consistent with a random location of their orbital paths are P= 5 $\times$ $10^{-8}$ and  0.0045 for the \citet{Bovy2015} and F18 models, respectively. The latter value is indeed consistent with the 3$\sigma$ dotted line that almost coincides with the F18 MW model. \\
However, one may object that we have accounted several times for the same orbital path, e.g., potential satellites of the LMC \citep{Erkal2019,Patel2020}. According to \citet{Patel2020}, there are four dSphs in the F18 sample that could be LMC satellites, Reticulum II, Horologium I, Hydrus I, and Carina III, assuming a very massive LMC. All of them lie near their pericenter as well as the LMC. To account for the effect of potential LMC satellites on our statistics, we exclude Reticulum II, Hydrus I, and Carina III from our sample, keeping Horologium I, which shows intermediate properties, in order to keep account for one LMC orbital path. It leads to 34 dSphs, for which the procedure described above  (nonparametric Kolmogorov)-Smirnov test) leads to a maximal distance of $D_{\rm max}$= 0.474 and an associated probability of 2 $\times$ $10^{-7}$. \\

It results that only a combination of high MW mass and a significant number of very small and undetected UFDs just below 300 kpc could be consistent with a random distribution of dSphs on their orbital paths. This is in strong tension with the rotation curve of the MW (see left panel of  Figure~\ref{fig:1}). It opens the possibility that dSphs are not satellites orbiting for a long time in the MW halo and are at their first infall. 

\subsection{Are most MW dSphs at their first infall?}

%However this could be well accommodated with the \citet{Bovy2015} MW-mass model if many dSphs were at their first passage. 
Hereafter we consider the same sample of 24 dSphs as in H19, which has been built from F18, and for which we have been able to describe scaling relations between the visible luminosity ($L_{\rm V}$), the half-light radius ($r_{\rm half}$), the velocity dispersion on the line of sight ($\sigma_{\rm los}$), and the distance to the MW ($D_{\rm MW}$). Three dSphs (Sagittarius, Crater II, and Bootes I) are clearly outliers in these relations, which leads to a sample of 21 dSphs (see H19 for more details).
The left panel of Figure~\ref{fig:2} shows that all (see open and filled magenta circles) but six dSphs (see black circles) have within 1 $\sigma$ their apocenters in excess of 245 kpc, which is the virial radius of the \citet{Bovy2015} model. Many (magenta) circles are between or above the two horizontal lines indicating 1-2 times the virial radius, an area that typically includes objects approaching the escape speed \citep{Deason2019}. If dSph progenitors are gas-rich dIrrs, the high apocenter values found by F18 are likely underestimated because ram pressure may have slowed down their motions when they arrived in the MW halo, and this until the gas has been fully extracted. This applies if the progenitors are DM-free, but it nevertheless leaves open the possibility that their orbits have initial eccentricities in excess of 1, and then for a first infall of most dSphs.\\

Although tangential motions provided by Gaia DR2 (F18) possess large errors, the right panel of Figure~\ref{fig:2} is a tentative way to reproduce the phase-space plot of \citet{Boylan-Kolchin2013} but for an MW mass model based on \citet{Bovy2015}.  Comparing it with Figure6 of \citet{Boylan-Kolchin2013}, it appears that most dSphs with large apocenters are consistent with a recent infall, $\le$ 4 Gyr ago (see points above or between the two magenta curves in the right panel of Figure~\ref{fig:2}). It also confirms that most dSphs are bound to the MW as shown by F18 (compare their locations to the escape velocity represented by the upper black dotted curve). However, being bound does not exclude a first passage, i.e., a first infall needs not, and almost always does not, imply an unbound orbit \citep{Boylan-Kolchin2013}. \\%Left panel of Figure~\ref{fig:2} shows that a large fraction of the 21 dSphs selected in \citet{Hammer2019} have very large apocenters, and that they could be consistent with a first passage. 
Having most of the dSphs at their first infall cannot be a full surprise since we already know that the Magellanic Clouds are at their first passage \citep{Kallivayalil2013} and that they are by far the largest contributors to the total mass of MW companions. One may also notice that most of the MW companions, including the LMC, have their locations and motions embedded in a Vast Polar Structure \citep[VPOS]{Pawlowski2014,Pawlowski2019,Fritz2018}, suggesting furthermore a common origin for them. We also note that this structure also includes five of the six dSphs (Carina, Draco, Fornax, Segue, UMi, but not Willman; see black circles in Figure~\ref{fig:2}) for which their orbits are fully confined within the virial radius within 1$\sigma$. One may notice that the orbit of Willman is not well constrained \citep{Pawlowski2019}, and it is the only object for which \citet{Simon2018} found a different orbit location (at pericenter) than that from F18 (at apocenter).\\
 The orbital properties and the exceptional concentration near their pericenters of the MW dSphs are not consistent with a classical scenario for which they are long-lived satellites and require a heavy DM subhalo to shield them from MW tidal forces. This leads us to investigate the tidal shock scenario for which MW dSphs are not at equilibrium, and hence without DM \citep{Yang2014,Hammer2015,Hammer2018,Hammer2019}.

\section{How do tidal shocks affect DM-free dSphs?}
\label{tdshocks}
\subsection{An exact calculation of the tidal shock effects}
Progenitors of Milky Way (MW) dwarf spheroidals (dSphs) are likely dwarf irregulars (dIrrs), from which gas has been stripped owing to the ram pressure caused by the Galactic halo gas \citep{Mayer2001}. All dwarfs (but a few, e.g., Cetus and Tucana) are gas-rich beyond 300 kpc and gas-poor within 300 kpc (except the LMC/SMC; \citealt{Grcevich2009}), which supports dIrrs being the dSph progenitors. The role of the gas during the process is essential, especially if dSphs are assumed to be DM-free. Gas removal by ram pressure of infalling dIrrs induces a lack of gravity implying that stars are then leaving the system following a spherical geometry. %Such a geometry ensures the dominance of tidal shocks over tidal stripping \citep{Binney2008}, i.e., 
The global instantaneous energy change $\Delta E$ caused by the MW tides on an individual star with velocity $\bf v$ is
\begin{equation}
 \Delta E = {\bf v \cdot \Delta v} + 1/2 (\Delta v)^2. 
 \label{Eq00}
 \end{equation}
The first term (called "tidal stripping" or "diffusion term"; see \citealt{Binney2008}, p. 663) vanishes when averaged over all stars, which
  explains the absence of tidally stripped features in many dSphs (H19), and this leads to $\Delta E = 1/2 (\Delta v)^2$.  If dSph kinematics are not affected by rotation \citep[see how velocity dispersion data are corrected for rotation]{Walker2009} the latter term (called "tidal shocking" or "heating") is approximated to $1/2 (\Delta \sigma^2)$, i.e., the kinetic energy increase due to tidal shocks.  In the frame of the impulse and distant-tide approximations (see a detailed discussion in H19), one may calculate the MW potential variations ($\nabla\Phi_{\rm MW} = - G M_{\rm MW}(D_{\rm MW})/D_{\rm MW}^2=g_{\rm MW}$ per unit mass, where G is the gravitational constant) within the dSph for small variations of the MW distance ($\Delta D_{\rm MW}$) under an assumed  spherical symmetry for the MW mass ($M_{\rm MW}$); one finds
\begin{equation}
\Delta \Phi_{\rm MW}  =  g_{\rm MW} \times \Delta D_{\rm MW},
\label{Eq01}
 \end{equation}
 
% and then:
 
%\begin{equation}
% \frac{\Delta \Phi_{\rm MW}}{\Delta D_{\rm MW}}=G M_{\rm MW}/D_{\rm MW}^{2} \times (1-\frac{\Delta M_{\rm MW}}{\Delta D_{\rm MW}} \times \frac{D_{\rm MW}}{M_{\rm MW}}),
%\label{Eq02}
% \end{equation}
% where the first and second term are the MW acceleration ($g_{\rm MW}$) and $\alpha_{\rm MW} = 1 - (\Delta M_{\rm MW}/\Delta D_{\rm MW})(D_{\rm MW}/M_{\rm MW})$ \citep{Hammer2019}, respectively. 
 Since velocity dispersion measurements are made at $r_{\rm half}$ and along the line of sight, one may consider the corresponding dSph volume that is a tube with radius $r_{\rm half}$ and elongated along the line of sight (Z-axis, see Figure 7 of \citealt{Hammer2018}). The variation of $\Delta \Phi_{\rm MW}$ can be estimated by subtracting the averaged potential exerted by the MW on stars at the farthest side of the tube from that exerted on  stars at the closest side to the MW \citep[see their Appendix B and Figure 7]{Hammer2018}.  To calculate $\Delta \Phi_{\rm MW}$ one needs to weight it with the dSph stellar density that is assumed to follow a Plummer profile, with $\rho = \rho_0 (1+r^2/r_{\rm half}^2)^{-5/2}=\rho_0/(4\sqrt{2}) \times (1+Z^2/(2r_{\rm half}^2))^{-5/2}$, assuming $r^2=Z^2+r_{\rm half}^2$. Accounting only for first-order variations in Eq.~\ref{Eq01}, it comes to weight $\Delta D_{\rm MW}= Z$ on each side of the dSph, finding	
\begin{equation}
 <\Delta D_{\rm MW}^{+}>=\frac{\int_{0}^{+\infty} Z \times \rho(Z) dZ}{\int _{0}^{+\infty} \rho(Z) dZ} =r_{\rm half}/\sqrt{2},
\label{Eq02}
 \end{equation}
and on the closest side, $<\Delta D_{\rm MW}^{-}>$ is calculated by integrating from $-\infty$ to 0 and is equal to $-r_{\rm half}/\sqrt{2}$. 
% \begin{equation}
% <\Delta D_{\rm MW}^{-}>=\frac{\int_{-\infty}^{0} \left( Z \right)  \left( 1+ Z^{2}/ \left( 2r_{\rm half}^{2} \right)  \right) ^{\frac{-5}{2}} dZ}{\int_{-\infty}^{0} ( 1+ Z^{2}/( 2r_{\rm half}^{2} ))^{\frac{-5}{2}} dZ}=-r_{\rm half}/\sqrt{2},
%\label{Eq04}
% \end{equation}
On both sides of the dSph, the condition $\Delta D_{\rm MW}$ $<<$ $D_{\rm MW}$ still holds because the density profile is very steep at the dSph outskirts. Accounting for the whole potential gradient between the two half tubes of stars leads to $\Delta \Phi_{\rm MW}= g_{\rm MW} \times 2r_{\rm half}/\sqrt{2} = \sqrt{2} \times g_{\rm MW} \times  r_{\rm half}$. %We verified that second order terms $\sim$ $\Delta D_{\rm MW}^2$ are cancelling each other, producing a null contribution. 
Assuming for all dSph stars an instantaneous exchange of energy in the impulse approximation ($1/2 (\Delta \sigma^2)=\Delta \Phi_{\rm MW}$), one finds 
 \begin{equation}
 \Delta \sigma^2 = 2\sqrt{2} \times g_{\rm MW} \times r_{\rm half} %=\sigma_{\rm MWshocks}^{2},
\label{Eq03}
 \end{equation}
where $g_{\rm MW}$ is the MW gravitational attraction that has to be (slightly) corrected to be projected on the line of sight, which is almost the direction made by the dSph and the Galactic center. In the following we will call the velocity dispersion increase ($\Delta \sigma$) brought by the MW tidal shocks, $\sigma_{\rm MWshocks}$. The main difference between Eq.~\ref{Eq03} and Eq. B16 of \citet{Hammer2018} is coming from the implicit assumption in the latter study that the MW potential applicable at the dSph distance is concentrated in a point mass, a similar assumption to that made by \citet[see their Eq. 4]{Aguilar1985}, but without adding a correcting factor. Eq.~\ref{Eq03} includes a full calculation of the tidal shock impact on a stellar population under the assumption of the impulse approximation, implying that the theoretical effect of tidal shocks has been strongly underestimated by \citet{Hammer2018}. \\

\begin{figure*}
	\includegraphics[width=1.8\columnwidth]{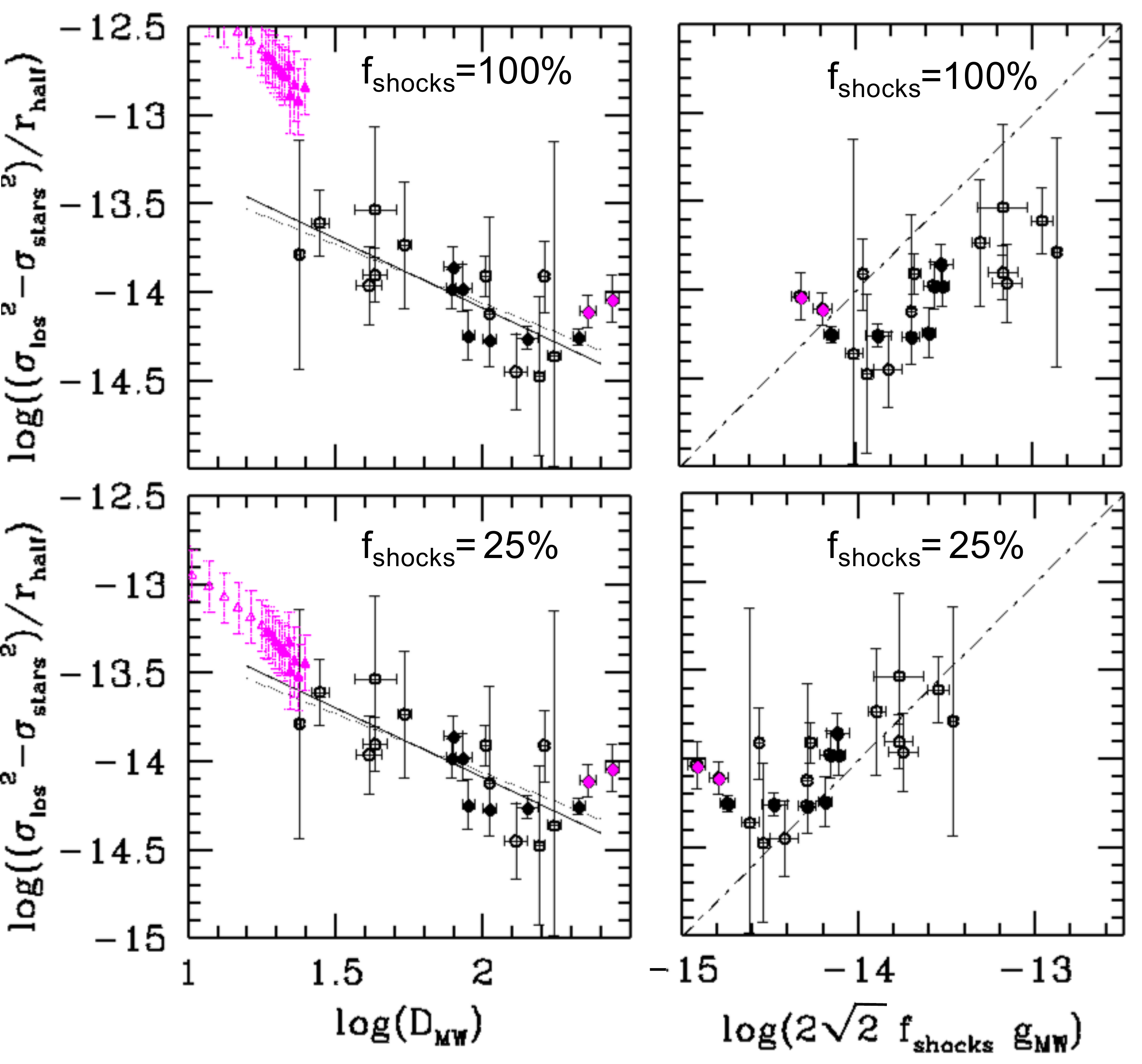}
    \caption{{\it Left panels:} tidal shocks (or DM) acceleration (in km $s^{-2}$) based on dSph kinematics (($\sigma_{\rm los}^2 - \sigma_{\rm stars}^2) \times r_{\rm half}^{-1}$) versus MW distance (in kpc). Data ($\sigma_{\rm los}$, $L_{\rm V}$, $r_{\rm half}$) are coming from the Table 1 of H19 (see references therein), with an update for the velocity dispersion of UMa II from \citet{Simon2019}, and for the Draco II luminosity from \citet{Longeard2018}. The figure compares the results from dSph data with those deduced through Eq.~\ref{Eq07} and based on the rotation curve from \citet[see filled magenta triangles with error bars]{Eilers2019}. Top and bottom panels shows how the rotation curve matches those of the nearest dSphs for $f_{\rm MWshocks}$ = 1 and 0.25, respectively. Filled and open circles represent classical and ultrafaint (nonclassical) dSphs, respectively, while magenta circles represent Leo I and Leo II, which do not obey to the impulse approximation (see text).  {\it Right panels}: same as the left panels, after replacing the MW distance in abscissa by the tidal shock acceleration (in km $s^{-2}$) predicted by the MW mass model from \citet{Bovy2015}, showing also a matching for $f_{\rm MWshocks} \approx 0.25$. %Variations of $M_{tot}$/L against the MW distance, with name indication of the dSphs having the same size than Crater.
    }
   \label{fig:3}
\end{figure*}

However, the impulse approximation is valid only if the encounter time is short compared to the crossing time, which is approximated to be $r_{\rm half}/\sigma_{\rm los}$ assuming that most or all dSph stars are in that regime (H19). This is unlikely because in most stellar systems the crossing time is a strong function of energy or mean orbital radius, so the impulse approximation is unlikely to hold for stars near the center. Indeed, sufficiently close to the center, the crossing times of most stars may be so short that their orbits deform adiabatically as the perturber approaches, and the encounter will leave most orbits in the central region unchanged \citep[see Sect. 8.2]{Binney2008}. This implies transforming Eq.~\ref{Eq03} into
 \begin{equation}
\Delta \sigma^2 =\sigma_{\rm MWshocks}^{2} = 2\sqrt{2} \times g_{\rm MW} \times r_{\rm half}  \times f_{\rm MWshocks},
\label{Eq04}
 \end{equation}
where $f_{\rm MWshocks}$ is the fraction of stars projected at $r_{\rm half}$, which obey to the impulse approximation\footnote{\citet{Weinberg1994} theorized the gravitational shocking that may occur even for slowly varying perturbations, i.e., beyond the impulse approximation, and then applicable to a wide variety of disturbances. It implies that $f_{\rm MWshocks}$ indeed represents the fraction of the system that is affected by tidal shocks.}. It may lead one to conclude that the effect is very difficult to observe, since this fraction may strongly vary from one object to another. This is unlikely to occur because the high-velocity motions due to tidal shocks for only a fraction of stars likely dominate systems with very small  self-gravity provided by their stellar content. The strong anticorrelation between the MW distance ($D_{\rm MW}$) and the acceleration caused by tidal shocks at $r_{\rm half}$ assuming $\sigma_{\rm MWshocks}^{2}=\sigma_{\rm los}^{2}-\sigma_{\rm stars}^{2}$ (see the left panels of Figure~\ref{fig:3} and also the left panel of Figure 1 of H19) confirms that the effect is sufficiently prominent to generate it in most dSphs.  \\

The precise MW velocity curve from Gaia DR2 \citep{Eilers2019} provides us a model-independent method to verify the MW gravitational attraction (and then tidal shocks) up to 20-25 kpc, i.e., close to the nearest dSphs. 
% and this reproduces quite precisely the observed dSph velocity dispersions as well as the fundamental relationships established from the observations (see Figs. 1-3 and 5-7 in \citealt{Hammer2019}).\\
%One may consider the above to be just coincidental, and that the anti-correlation can be explained by a still unknown process. 
MW disk stars obey to the virial theorem equation, $V_{\rm rot}^2 = G\times M_{\rm MW}(D_{\rm MW}) / D_{\rm MW}$, and then
\begin{equation}
g_{MW}= G\times M_{\rm MW}(D_{\rm MW}) / D_{\rm MW}^2 = V_{\rm rot}^2 / D_{\rm MW}.
\label{{Eq05}}
 \end{equation}
One may deduce from
\begin{equation}
a_{\rm MWshocks} = (\sigma_{\rm los}^2 - \sigma_{\rm stars}^2)/r_{\rm half} =2\sqrt{2} \times g_{\rm MW} \times f_{\rm MWshocks},
\label{Eq06}
 \end{equation} 
 that the MW velocity curve may predict the tidal shock acceleration:
\begin{equation}
a_{\rm MWshocks} = 2 \sqrt{2} \times V_{\rm rot}^{2}/D_{\rm MW} \times f_{\rm MWshocks}.
\label{Eq07}
 \end{equation}
The bottom panel of Figure~\ref{fig:3} shows that for $f_{\rm MWshocks} \approx  0.25$, the MW velocity curve predicts a tidal shock acceleration at 20-25 kpc that matches well with that derived from the nearest dSphs. It is only the unprecedented accuracy of the MW velocity curve that allows such a test. Right panels of Figure~\ref{fig:3} illustrate that for the same $f_{\rm MWshocks} (\approx 0.25$), the \citet{Bovy2015} MW mass profile provides a good agreement between the tidal shock acceleration ($a_{\rm MWshocks}= 2\sqrt{2} \times g_{\rm MW} \times f_{\rm MWshocks}$) and its measurements from the properties of all dSphs  (($\sigma_{\rm los}^2 - \sigma_{\rm stars}^2)/r_{\rm half}$). It means that  in most dSphs there are enough stars affected by the MW tidal shocks to generate the strong correlations shown in Figure~\ref{fig:3}. \\
%If the MW possessed a flat rotation curve ($\Delta V_{\rm rot}/\Delta D_{\rm MW}$= 0) from 6 to 20 kpc, it would have led to a null value for the tidal shock acceleration up to 20-25 kpc (i.e., a null value for $\alpha_{\rm MW}$, see Eq.~\ref{Eq07}). Then, in the tidal shock scenario, the kinematics of Segue suffices to predict the slope of the MW velocity curve at $\sim20-28$ kpc within the \citet{Eilers2019} error bars.\\

\subsection{Tidal shocks in pure N-body simulations}
\label{simus}
Several studies have considered the effects of tidal stripping and shocks on dwarf satellite galaxies, including DM-free dSph progenitors \citep{Piatek1995,Kroupa1997,Read2006,Fellhauer2008,Smith2013}. However, they do not describe the combination of physical processes that reproduce dSphs. The four first studies have not considered the presence of the gas whose removal induces the dominance of tidal shocks (see Eq.~\ref{Eq00}), leading to strongly tidally stripped galaxies, whose elongations along the line of sight are far too large to be consistent with the observed horizontal branch \citep{Klessen2003,Read2006}. The latter study only accounts for the gas removal by ram pressure using a "wind-tunnel" model. It shows that gas removal may help to increase the velocity dispersion, although it does not account for the unavoidable tidal forces exerted by the MW. \\

\citet{Yang2014} were the first to fully account for gas removal and tidal shocks using numerical simulations with the GADGET2 code \citep{Springel2005}. However, understanding stellar systems out of equilibrium requires particle masses consistent with those of individual stars  to avoid non-Newtonian approximations introduced by softening. For doing this we have chosen to use pure N-body simulations with only star particles (masses from 0.1 to 1 $M_{\odot}$), for dSph progenitors with initial stellar masses from $10^4$ to 2.5 $\times$ $10^6 M_{\odot}$. The infalling dSph progenitors are assumed to be gas-rich objects with gas fractions from 50 to 90\%. Density profiles for gas and stellar components are described by a modified Hubble profile \citep[see their Eq. 2.52]{Binney2008}. Simulations were performed using software (GIZMO; \citealt{Hopkins2015}), under a similar physical frame to that in \citet{Yang2014}, but using the MW mass model by \citet{Bovy2015}, and higher gas density for the MW gas, which has been modeled by \citet[see their model 28]{Wang2019} to successfully reproduce both the HI and HII components of the Magellanic Stream. This allows us to represent the MW gravitational field by an analytic MW potential, as well as to resolve analytically the interaction between the dwarf gas and the MW halo gas. The gas removal from dwarf galaxies is analytically calculated using the \citet{Close2013} model, for which two physical processes are considered, i.e., the balance between the gravitational force of the dwarf galaxy and the external ram pressure (i.e, from the \citealt{Gunn1972} formula), and the mass loss introduced by Kelvin-Helmholtz instabilities (e.g., \citealt{Nulsen1982}).  We have verified that the above technique of "pure N-body for stars plus analytical ram pressure for gas" is able to reproduce the  gas removal process resolved by hydrodynamics simulations done in \citet{Yang2014}, and in \citet{Hammer2018,Hammer2019}. \\ %And the gain is that there is no more suspicious on stellar dynamics in dwarf galaxies because particles are now represent real stars.\\

\begin{figure*}
\centering
\includegraphics[width=1.95\columnwidth]{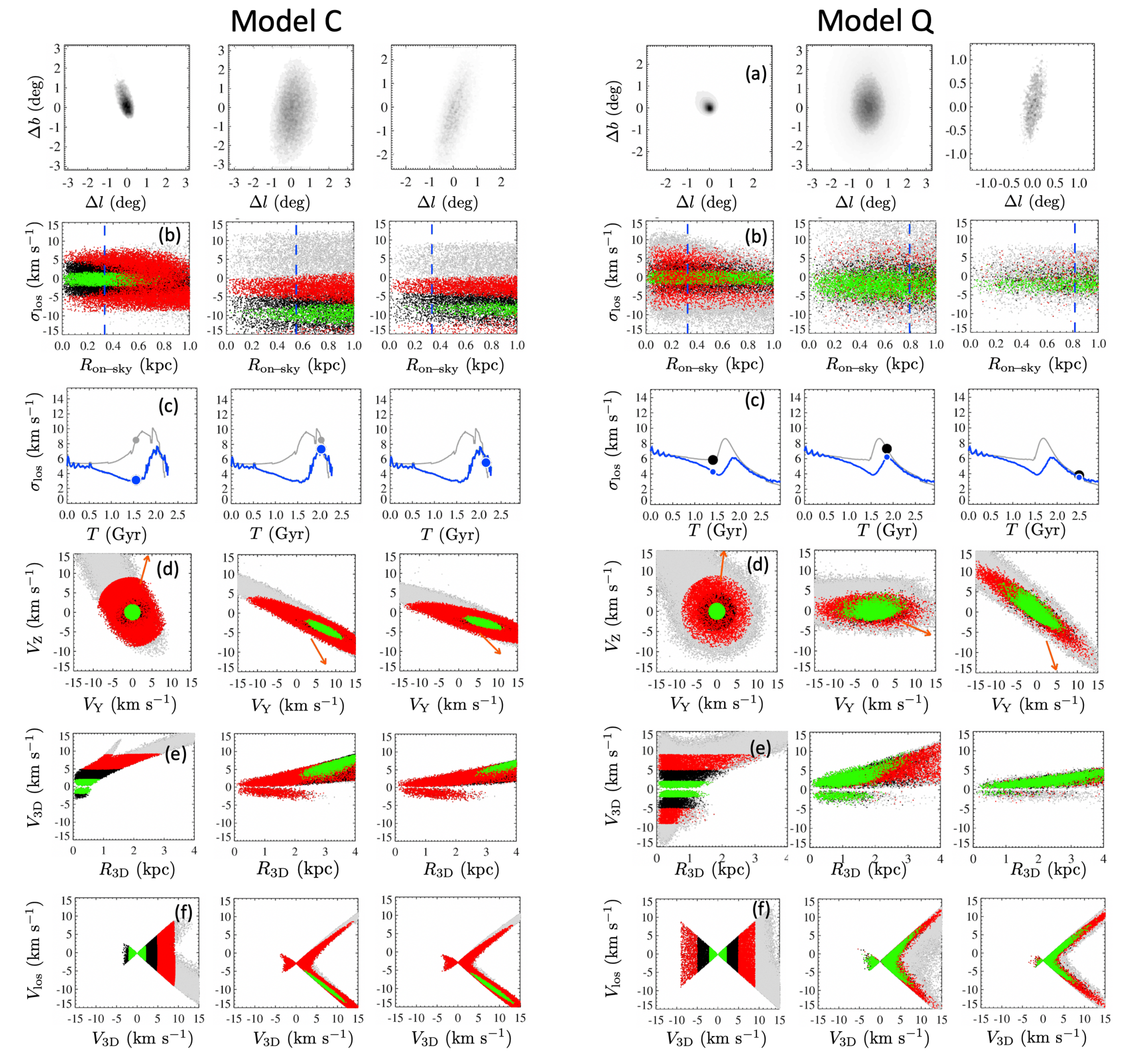}
    \caption{Summary of two N-body simulations of a DM-free and gas-rich dSph progenitor, with three columns on the left (right) for model C (model Q), respectively. They are shown at steps 1, 2, and 3, from left to right, respectively (see text). The first line shows the appearance of the simulated galaxies in gray colors. For other lines, green, red, black, and gray symbols represent stars in step (1), from very low velocity and very near the center to very high velocity and very far from the center (gray symbols with extremely low density); see left panels of the (d)-(f) rows. {\it From top to bottom:} (a) projected view on sky; (b) $V_{\rm los}$ versus sky projected radius, with the vertical blue dashed line showing $r_{\rm half}$ that includes the selected stars; (c) the evolution of $\sigma_{\rm los}$ with time showing a peak very near the pericenter passage (see point in middle panel), and the gray curve indicating the theoretical expectation from Eq.~\ref{Eq04} for $f_{\rm MWshocks}$ = 0.25 (0.08) for model C (model Q), which also includes self-gravity from gas and stars; (d) the velocity distribution projected in gal-(Y, Z) plane in Galactocentric coordinates, and with the arrow showing the MW direction; (e) the phase diagram indicating that many stars are leaving the system after gas removal (left panel) ; and (f) $V_{\rm los}$ versus $V_{\rm 3D}$ showing that near pericenter (middle panel) most stars have their motions only on the line of sight.
     }
    \label{fig:4}
\end{figure*}

The above allows us to probe the interaction of dSph stars with the MW gravitational field in a full Newtonian scheme. By adopting a gravitational force softening of 0.005 pc, it implies that there should be no stellar particles that can be affected by the undesirable effects due to the softening in media with central stellar density $\sim$ 500 times smaller than at the Sun neighborhood (see H19).  From left to right, Figure~\ref{fig:4} shows three different steps in evolution of such a stellar system after the gas removal, i.e., when it may resemble a dSph. It includes (1) when  the gas has just been fully removed, (2) just after pericenter at the peak of tidal shock effects, and (3) well after pericenter, when the system almost vanishes due to the heavy tidal stripping process that unavoidably follows tidal shocks.\\
 In Figure~\ref{fig:4}, the initial dSph progenitors have a stellar mass of 1.9 $\times$ $10^6 M_{\odot}$, $r_{\rm half}$= 230 pc,  and a gas fraction of 95\% (model C, for model Q the values are 8.1 $\times$ $10^6 M_{\odot}$, 440 pc, and 90\%, respectively). Progenitors are initially set at 340 kpc away from MW center, on an orbit with an eccentricity of 1.2 and a pericenter of 50 kpc. In step 1, stars (green, red, and black symbols in the left panels of Figure~\ref{fig:4}) distribute in a roundish geometry, and their velocities (panels (d)-(f)) show no preferential direction along the line of sight or toward any direction. This corroborates the theoretical prediction that the MW tidal shocks will become the dominant tidal process for a spherical geometry (see Eq.~\ref{Eq00}). Middle panels of Figure~\ref{fig:4} show that just after pericenter, at step 2, most stars are in resonance with the MW gravitational forces, which force stars to align their velocities near the line of sight (see panels (d) and (f)). Later on the object is sufficiently elongated by tides that tidal stripping becomes the dominant effect, destroying rapidly the dSph, and leaving a faint and elongated residue that resembles to a stream (see panel (a)). This could explain the difficulty of observing dSphs far from their pericenters.\\
  We have attempted several tests using different mass and kinematics profiles for the dSph progenitors, which will be described in a future paper (Yang, Y. B. et al. 2020, in preparation). The simulations presented here are still limited as they show gas removal near 50 kpc (at pericenter), while it is expected to occur at 200-300 kpc according to \citet{Grcevich2009}, since there are almost no dwarfs with gas within 300 kpc. One may try to detect the stellar particles (see gray points in panels (d)-(f) of Figure~\ref{fig:4}) attracted by the gas during its removal by ram pressure, providing a potential test of the DM-free dSph scenario (see also Figure 1 of \citealt{Smith2013}, and \citealt{Yang2014}). This could be attempted for the few objects having lost  their gas recently. Another limitation is the extremely low surface brightness expected for this phenomenon, which is several magnitudes fainter than the lowest isophotal levels of dSphs. We also find that tidal shocks are more efficient for a modified Hubble profile than for a Plummer body, which is not unexpected since the latter are known to be more robust against shocks as they are hypervirial bodies \citep{Evans2005}.

\section{Can properties of all dSphs be explained by MW tides?}
\label{All}
Here we reexamine the objects classified in F18 and that have been excluded by H19 because of their disputed classification (e.g., Crater), because of their proximity to the Magellanic Clouds (e.g., Carina II, Reticulum II, and Hydrus), and finally since only an upper limit to their velocity dispersion has been determined (e.g., Tucana III, Triangulum II, Segue 2, Grus I, and Hydra II). As in H19, we only consider objects having eight or more stars for establishing their kinematics.

\subsection{Reinserting Crater into the dSph sample}
\label{Crater}

\begin{figure}
	\includegraphics[width=1.0\columnwidth]{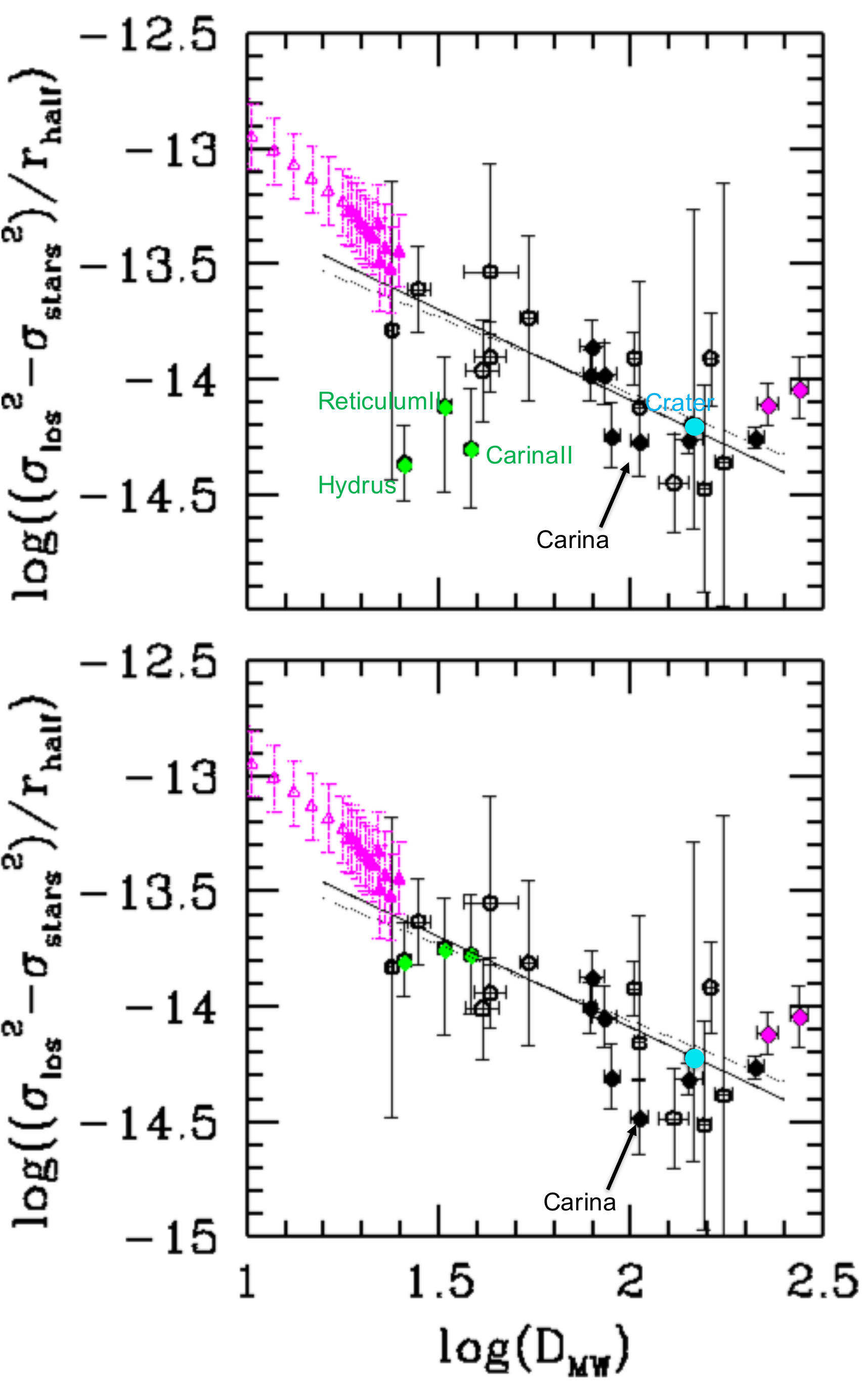}
    \caption{{\it Top:} same as the bottom left panel of Figure~\ref{fig:3} in which we have introduced the three dSphs that lie in between LMC and the Galactic Center (green points; see Sect.~\ref{MS}) and Crater (cyan point; see Sect.~\ref{Crater}).  {\it Bottom:} same as the top panel, but for which we have applied to all dSphs the effect of the Magellanic System assumed to be concentrated at the LMC location, i.e., the ordinate has been subtracted with $a_{\rm MSshocks} \times \cos(\Psi_{\rm LMC})$ (see Eq.~\ref{Eq08}). Besides the three dSphs (green points) that move up to reach the anticorrelation, an arrow indicates Carina because its location further than the LMC leads to move the corresponding point toward the opposite direction. Data for the three additional dSphs are coming from \citet{Mutlu-Pakdil2018} and \citet{Minor2019} for Reticulum II, from \citet{Koposov2018} for Hydrus, and from \citet{Torrealba2018} for Carina II. Data for Crater are coming from \citet{Weisz2016}. }
    \label{fig:5}
\end{figure}

Crater is an intriguing stellar system, originally
discovered independently by \citet{Belokurov2014} and
by \citet{Laevens2014}.
Right from the beginning it was unclear whether this object
should be classified as a cluster or a dwarf galaxy. 
The main observational evidence that suggests
that Crater is a galaxy is the existence of an extended "blue plume''
that can be interpreted as a younger population \citep{Belokurov2014}. 
There have been spectroscopic measurements of Crater stars by
\citet{Bonifacio2015}, \citet{Kirby2015}, and \citet{Voggel2016}.
Of these, only \citet{Bonifacio2015} provided detailed chemical
abundances for the two observed stars (12 different elements). 
The only chemical characteristic that may point
toward a globular cluster (GC) would be the presence
of an Na-O anticorrelation \citep{Gratton2001,Bastian2018,Carretta2019}.
However, oxygen is not measured in any of the two stars, and sodium
provides [Na/Fe] $\sim -0.25$ for both stars.
\citet{Weisz2016} obtained deep HST photometry for Crater and
revisited its color-magnitude diagram (CMD). 
From their analysis they concluded
that Crater is a GC, because their fit to the the observed 
CMD with a single stellar population was better than that with a
variable star-formation history. 
However, note that, in both cases, they failed to fit the observed
"blue plume'' and concluded that it must consist of
blue stragglers. As pointed
out by \citet{Bonifacio2015} the star clusters (both globular and
open) define a rather tight anti-correlation between the blue straggler
frequency\footnote{defined as the logarithm of the number
of blue stragglers divided by the number of horizontal branch stars:
log($N_{BSS}/N_{HB}$)} and the absolute luminosity of the cluster
\citep{Momany2015}. In practice, the more luminous the cluster, the fewer
the blue stragglers. In this respect the number of observed blue stragglers
in Crater is too large with respect to globular clusters
of the same absolute luminosity, and instead similar
to what is observed in dwarf galaxies of similar luminosity
like Leo IV or Bootes. So if Crater were a GC, it would be
a very peculiar GC.
The nature of Crater, GC or galaxy, is still open to debate, 
since we lack any clear-cut way to decide.
Probably the most promising route is the measurement
of Na abundances in a larger sample of stars, to check
for the existence of a sizable dispersion, which would
unequivocally signal a GC.\\
\citet{Voggel2016} measured radial velocities for 26 members of Crater
and derived $\sigma_{\rm los} = 2.04^{+2.19}_{-1.06}$ $km s^{\rm -1}$. This leads to a low $M_{tot}/L_v$ value just below 10, which left Crater at odds with expectations for such a tiny galaxy, e.g., it contrasts a lot with values near or above 1000 for Draco II and Segue, which have a similar small radii to Crater, i.e., close to 20 pc.
From the assumption that a galaxy of that luminosity must be very DM dominated, \citet{Voggel2016} concluded that Crater is a GC. They also discussed the possibility that the velocity dispersion is caused by tidal disturbances letting the object out of equilibrium. \\%Note that if we relax the above assumption, the conclusion does not follow.
Figure~\ref{fig:5} shows that most properties of Crater (see the cyan point) are consistent with this possibility, and that it is indeed sharing the same correlation as dSphs affected by tidal shocks.  %Draco II, Segue and Wilman I have been also singularized in Figure~\ref{fig:2} (see cyan points),  % In the DM scenario, the low $M_{tot}/L_v$ value for Crater is at odds with expectations for such a tiny galaxy, e.g., it is around a thousand for  Draco II, Segue and Willman I. 
Tidal shocks are far much less efficient far from the MW, which explains the discrepancy between $M_{\rm tot}/L_{\rm V}$ between Crater (less than 10) and the two nearest dSphs (close to 1000 for Segue and Draco II; see the black open circles on the left of Figure~\ref{fig:5}). The three objects have the same size, and tidal shocks explain well the $M_{\rm tot}/L_{\rm V}$ since it depends only on the MW distance or gravitational attraction. One may predict that all tiny systems lying much farther than few tens of kiloparsecs should have small $M_{\rm tot}/L_{\rm V}$ values (see the location of Crater in the bottom panel of Figure~\ref{fig:5}), and perhaps some have not been entirely discovered because of their size, distance, and faintness.

\subsection{The influence of the Magellanic System on Carina II, Reticulum II, and Hydrus}
\label{MS}
These three dSphs are located just in between the Galactic center and the LMC, i.e., they can be affected by tidal shocks from both systems. It leads to reducing the potential variation within these dSphs (see Eq.~\ref{Eq02}), i.e., the acceleration due to LMC (or Magellanic System) tidal shocks, $a_{\rm MSshocks}$ is in the opposite direction to that due to MW tidal shocks, $a_{\rm MWshocks}$. This is in good agreement with the fact that these galaxies  lie below the relationship made by the sample of 21 dSphs (see the top panel of Figure~\ref{fig:5}). This also explains why they show small $M_{\rm tot}/L_{\rm V}$, e.g., \citet{Koposov2018} found that Hydrus has $M_{\rm tot}/L_{\rm V}$ = 66, which is significantly smaller than typical values for UFDs at similar luminosity.  \\
The Magellanic System (MS) mass distribution is likely complex since the Magellanic Clouds are in strong interaction and they have lost significant fractions of their initial gas in the Galactic halo \citep{Fox2014} probably through ram pressure effects \citep{Hammer2015,Wang2019}. Assuming that all the Magellanic System mass is concentrated at the LMC location allows us to verify whether it may affect the dSph kinematics. To correct the MS effect, one may replace the ordinate of Figure~\ref{fig:5} by

 \begin{equation}
% \Delta \sigma^2 = 2\times \sqrt{2} \times g_{\rm MW} \times \alpha_{\rm MW} \times r_{\rm half} =\sigma_{\rm MWshocks}^{2},
\frac{\sigma_{\rm los}^2 - \sigma_{\rm stars}^2}{r_{\rm half}} - a_{\rm MSshocks} \times \cos(\Psi_{\rm LMC}) ,
\label{Eq08}
 \end{equation}
where $a_{\rm LMCshocks}= 2 \sqrt{2} \times GM_{\rm MS}/r_{\rm LMC}^2$ and $\Psi_{\rm LMC}$ is the angle between the line of sight and the direction made by the dSph and the LMC. The bottom panel of Figure~\ref{fig:5} shows how the three dSphs (see green points) are reaching the relation made by other dSphs when accounting for an MS mass of $M_{\rm MS}$= 2	$\times$ $10^{10} M_{\odot}$ at the LMC location.  Even if the above is based on gross assumptions, the derived mass value is not unrealistic in being not considerably higher than the sum of the gas and stellar mass associated with the MS, including the Magellanic Clouds. Assessing more accurately the MS mass and its spatial distribution requires a far better modeling and should also account for other MW dSphs, e.g., Carina (see the arrow in Figure~\ref{fig:5}), which is affected in the opposite direction since it lies beyond the LMC.

\subsection{A tentative classification scheme for the MW dSphs}
Figure~\ref{fig:6} summarizes the properties of all MW dSphs for which kinematics has been sampled (F18, see their Table 1). It leads us to suggest a new classification scheme that is based on the coherence between their orbital and their internal kinematics properties. Two broad categories can be distinguished:
\begin{figure}
	\includegraphics[width=1.0\columnwidth]{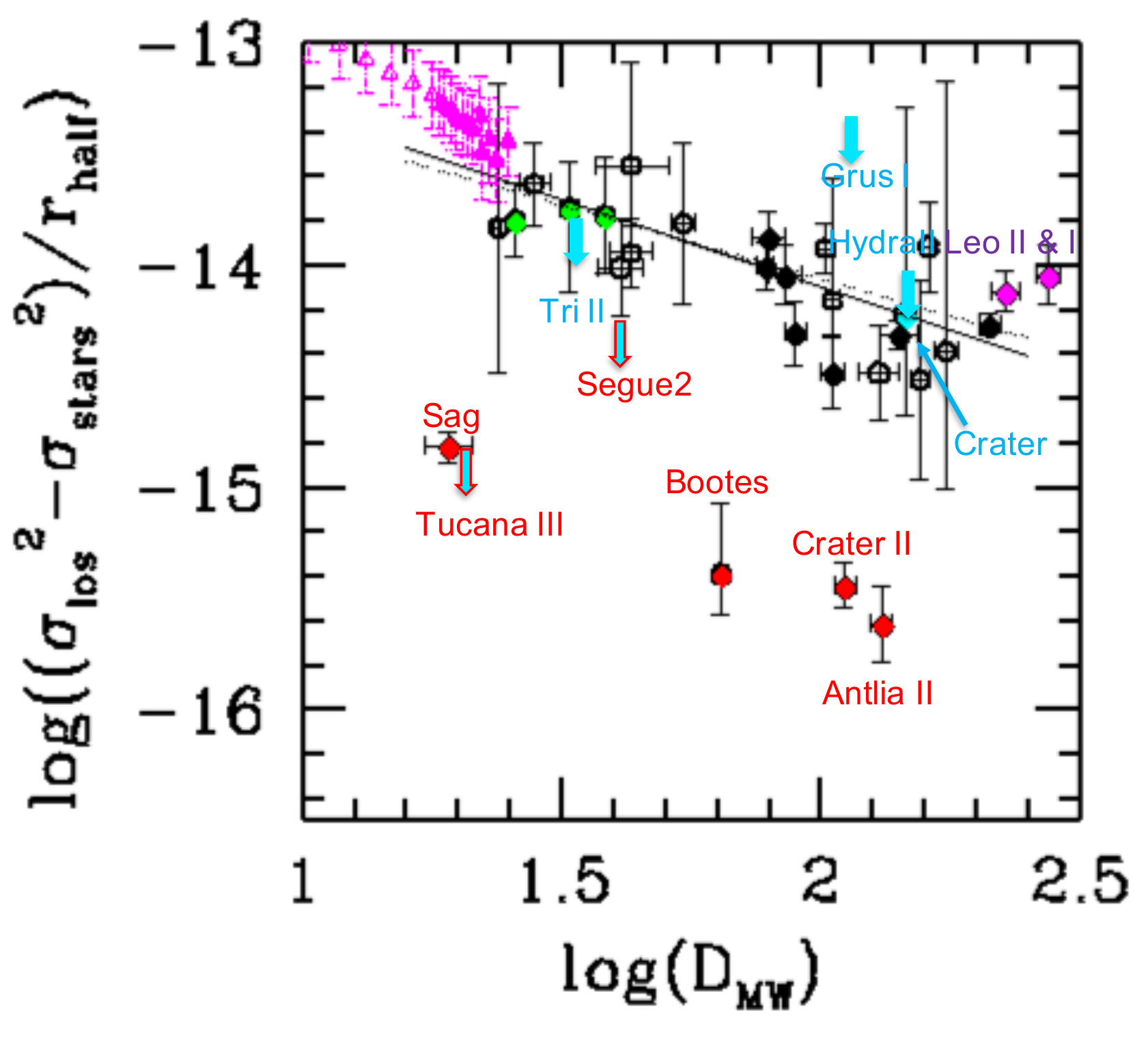}
    \caption{Summary of the properties of all MW dSphs for which kinematics has been sampled by 8 stars or more. Filled (open) black circles indicate classical (nonclassical) dSphs. Leo I and Leo II (magenta points) have been singularized because they do not obey to the impulse approximation, as well as the three dSphs in between LMC and the Galactic center (see green circles). Red circles show dSphs for which the tidal shock acceleration is smaller by 1 dex when compared to the anticorrelation, i.e., the tidally stripped dSphs after more than one passage. Data for Sag, Crater II, and Bootes are from Table 1 of H19, and data for Antlia are from \citet{Torrealba2019}. Vertical arrows indicate upper limits for Tucana III (data from \citealt{Simon2017}), Triangulum II \citep{Kirby2017}, Segue 2 \citep{Kirby2013}, Grus \citep{Walker2016}, and Hydra II \citep{Kirby2015}.}
    \label{fig:6}
\end{figure}

\begin{itemize}
\item The "tidally shocked dSph sample" includes all galaxies that follow the anticorrelation between $(\sigma_{\rm los}^2 - \sigma_{\rm stars}^2)/r_{\rm half}$ and the MW distance. Since the anticorrelation naturally appears from the MW tidal shocks, these objects are likely at first infall, generally near or after their passage to the pericenter. The 21 dSphs from H19 are part of this sample, and their orbits are either eccentric (see Figure~\ref{fig:2}), or they are part of the VPOS (see Sect.~\ref{orbits}) that includes the very eccentric LMC orbit. The impulse approximation is, however, not verified by Leo I and Leo II (shown in magenta in Figure~\ref{fig:6}), which are also discrepant to the anticorrelation. Leo I velocity dispersion is probably affected by tidal stripping since its motion is almost parallel to the line of sight, and accounting for this would suffice to bring it on the anticorrelation of Figure~\ref{fig:6}. To this category, one may add Crater (cyan point; see Sect.~\ref{Crater}), and also Hydrus, Reticulum II, and Carina II (green points) after accounting for the tidal forces of the Magellanic system (see Sect.~\ref{MS}). Within the uncertainties provided by the upper limit to their velocity dispersion, Grus I and Hydra II could belong to the same category. Note that all the six additional objects have an apocenter well in excess of 245 kpc (see Figure~\ref{fig:2}), except Reticulum II (see F18). The "tidally shocked dSph sample" represents the bulk of MW dSphs  with 26 objects. 
\item The discovery of Sagittarius \citep{Ibata2003} and its stream demonstrates that some dSphs may have experienced several orbits before being fully destroyed because tidal stripping becomes increasingly efficient after several passages \citep{Piatek1995,Kroupa1997}\footnote{The \citet{Kroupa1997} solutions are obtained by starting the satellites in dynamical equilibrium and allowing the repeated tides from the MW to remove stars until a stable remnant is obtained.}. From their locations in Figure~\ref{fig:6} (see red points) the "tidally stripped dSphs sample" includes Sag, Bootes I, and Crater II, which have been often considered to be anomalous objects and that do not obey to the scale relations between radius, stellar mass, velocity, and MW distance defined by the bulk sample (H19, see their Appendix A). It is likely that Tucana III is also part of the "tidally stripped dSph sample" since its orbit is almost purely radial with a pericenter of only 3 kpc (F18). Tucana III has been almost fully destroyed after its passage that close to the Galactic center, and it has now passed its apocenter. Note that Tucana III has an orbit fully enclosed into the MW virial radius, as well as Sag, Bootes I, and Crater II, which is compliant with their kinematics properties.
\end{itemize}
Figure~\ref{fig:6} also includes the recently discovered Antlia II \citep{Torrealba2019} that is not part of the F18 sample but has very similar properties to Crater II. Only the Triangulum II and Segue II (and possibly Grus I and Hydra II) locations in Figure~\ref{fig:6} are still ambiguous to the above-proposed classification.

\section{Discussion}
\label{discussion}
Literature estimates of the DM content in dSphs have been derived from their high velocity dispersions along the line of sight ($\sigma_{\rm los}$). These estimates assume that these systems are in  self-equilibrium and use a relation that is identical\footnote{H19 has shown that the calculation of the dSph DM content \citep{Walker2009,Wolf2010} comes from the measurement of the self-gravity attraction projected along the line of sight, which is $a_{\rm DM}= G M_{\rm DM} \times r_{\rm half}^{-2} = (\sigma_{\rm los}^2 - \sigma_{\rm stars}^2) \times r_{\rm half}^{-1}$. The latter quantity is also the theoretical expectation for the acceleration caused by MW tidal shocks at the half-light radius of DM-free dSphs (see Eqs.~\ref{Eq04} and ~\ref{Eq06}).} to that derived from DM-free objects dominated by tidal shocks (H19). Why do MW tidal shocks predict that the DM mass-to-light ratio of Segue is around 1000, while that of Fornax or Crater is only 10? Is it possible that the inference that the high velocity is due to the presence of DM is a misinterpretation?  The gravitational attraction or acceleration  ($a_{\rm DM}$) caused by the DM strongly anticorrelates with the dSph distance from the MW (see the top panel of Figure~\ref{fig:1}), which is still not explained if dSphs are DM dominated. H19 (see their Figure 8) showed that for DM-dominated dwarfs the DM density may decrease with a MW distance increase as it could be expected if tidal stripping is at work. However, this appears at odds with the star density within the same radius, which is independent of the MW distance. To overcome this, we would need an explanation of why tidal stripping is so discriminative between DM and stars (see H19). The answer on which scenario is valid for dSphs, DM-dominated or DM-free, relies on (1) the credibility of the tidal shock scenario that has been recently proposed and verified here by theoretical calculations and N-body simulations and (2) the observational facts that can be interpreted by one scenario but not by the other.\\

The orbital motions of the classical dSphs show a large eccentricity range, while most of the orbits of the nonclassical dSphs (UFDs) have been revealed by Gaia DR2 to be very eccentric. A recent infall is consistent with most dSph locations in Figure~\ref{fig:2}, including in the phase-space diagram (see magenta points) in agreement with the findings of \citet{Boylan-Kolchin2013}. Could this be consistent with having most dSphs at their first infall? A negative answer would suffice to falsify the tidal shock scenario, since it is likely that tidal stripping effects become dominant after more than one orbit. There are few tidally shocked dSphs that appear not consistent with a first infall in Figure~\ref{fig:2}, since they have small orbital energy and eccentricity. This includes Segue, Willman, UMi, Draco, Carina, and Fornax (see black points in Figure~\ref{fig:2}). The most massive ones, Fornax and Carina, were forming stars \citep{de Boer2012,Weisz2016} less than 1 and 3 Gyr ago, respectively. This unavoidably indicates the presence of gas at these epochs, and then a relatively recent infall. \\

The reason why some dSphs appear to have no star younger than 8-10 Gyr may appear at odds with a recent infall. %Comparison to Rocha (2012). This has been done following \citet{Fillingham2019} by "normalizing" the binding energy of each satellite by dividing the measured binding energy by the present-day host maximum circular velocity squared ($V^{2}_{\rm max}$, defined in Eq. 6 of \citealt{Bullock2001})).
In the frame of the tidal shock scenario, the gas removal process during the infall of dSph progenitors could profoundly affect the final dSph properties, including their orbits and their SFHs. For example, ram pressure effects have not been considered by \citet{Boylan-Kolchin2013}, while in the current simulations shown in Figure~\ref{fig:4}, they cause a slowdown, changing the eccentricity from an initial 1.2 to 0.7 at pericenter (see also \citealt{Yang2014}). It left few doubts that UFDs with large eccentricities are at their first infall. The slowdown due to ram pressure effects may also explain the few dSphs showing low eccentricities. Even SFHs can also be profoundly affected by the gas removal process. For example, during the infall one may expect stars to be formed from the pressurized gas. It is possible that these young stars could have very different kinematics than that of older stars. Such a differentiation may help them to escape the system when it inflates just after the gas removal and the subsequent loss of gravity.
%Moreover, the most recently formed stars (e.g., during the phases during which the dwarf gas is pressurized by the MW gas) may have higher velocities due to their origin from a turbulent medium, and can be later removed from the dSph after the gas exhaustion and the subsequent loss of gravity.
%it is expected that the ram pressure process is less efficient in removing the gas in objects with small eccentricity orbits, which likely implies a longer and more efficient slowdown of the orbital motion, i.e., a decrease of the orbital energy and then of apocenter and eccentricity. 
Such processes have to be modeled and simulated, and this will be the purpose of a future paper (Yang et al. 2020, in preparation). In the meantime, no robust conclusions can be made from the dSph SFHs. Same caution should apply as well for the DM content of the dSph progenitors, which are likely dIrrs, in a range of small stellar masses for which rotation curves are not accurate (see H19 for a full discussion, and also recent results from \citealt{Guo2019}). \\

However, the most intriguing property of dSphs is the fact that most of them are found near their pericenter \citep{Fritz2018,Simon2018,Simon2019}, for an MW mass model consistent with its accurate rotation curve \citep{Eilers2019,Mroz2019}, with both results coming from Gaia DR2 studies. \citet{Cautun2019} have attempted to supersede the \citet{Bovy2015} MW mass model by using a contracted DM halo, but it does not fit the most external part of the MW rotation curve and increase by only 20\% the MW total mass. Beyond a doubt, the large fraction of dSphs near pericenter excludes the possibility that dSphs are long-lived satellites of the MW (P$\sim$ 2 $\times$ $10^{-7}$ according to Kolmogorov-Smirnov statistics; see Sect.~\ref{orbits}).  It relaxes the need of self-equilibrium in most dSphs, and then for DM, while its other pillar, the prediction of the high velocity dispersion in dSphs, is also reproduced by MW tidal shocks, in absence of DM (H19, see their Figure 3). The latter scenario is fully consistent with the dSph locations near their pericenters,  where their structures are preserved for a short time by tidal shocks through the resonance of their stars with the MW tidal field (see Figure~\ref{fig:4}, Sect.~\ref{simus} and below). 

In summary, passages near pericenter are quite disruptive, which is expected and could lead to the full destruction of the dwarf galaxy. A full set of simulations is needed to  estimate the tidal shock impact that depends on the pericenter value, the mass of the progenitor, the initial gas fraction, and the orbit eccentricity. This suggests a relative paucity of dSphs leaving to their apocenter after a disruptive passage to their pericenter. However, this does not apply for galaxies approaching their pericenter for the first time, and one would expect to observe dIrrs or recently gas-stripped dSphs well before their pericenter. Fornax is one of them, perhaps one of the last newcomers, since it has lost its gas only $10^{8}$ yr ago \citep{Coleman2008,Battaglia2012} and is approaching the MW center. An overall picture is needed, which may imply that tidally shocked dSphs are arriving in a quite ordered way, the last one being Fornax, the first ones being generally destroyed by their passage to pericenter. Such a possibility had been discussed by \citet{D'Onghia2008} for interpreting the VPOS \citep{Pawlowski2014}) that includes the LMC. Later on, \citet{Fouquet2012} and \citet{Hammer2013,Hammer2015} have considered a structured arrival of tidal dwarfs, for interpreting together the occurrence of the VPOS, the similar structure surrounding M31 \citep{Ibata2013}, the Magellanic Stream, and the Leading Arm.

% \citet{Boylan-Kolchin2013} indeed described Leo I as being at its first pericentric passage, and from its large and positive radial velocity, it could not have reached its first apocenter yet. Its first infall occurred 1.5-2 Gyr ago, a time consistent with the star formation history \citep{Weisz2016}.

\section{Conclusion}
The past history of dSphs, including the nature of their progenitors and the age of their stars, is not sufficiently compelling to distinguish whether most dSphs are DM dominated at equilibrium or DM-free tidally shocked. 
 However, the present knowledge of their past orbits allows us to compare the ability of the two scenarios to reproduce the dSph properties. The tidal shock scenario is compelling because it:
 \begin{itemize}
 \item fully explains why an excessive fraction of dSphs is found near their pericenter, while the possibility that dSphs are long-lived satellites is excluded with a probability of P$\sim$ 2 $\times$ $10^{-7}$, which is at odds with the DM-dominated dSph scenario;
 \item predicts the significant anticorrelation with a probability to be fortuitous of 3 $\times$ $10^{-4}$ (see H19) between the acceleration associated with either tidal shocks or to DM and the MW distance, while such a relation seems not consistent with the DM scenario;
 %\item is predicted in amplitude (fraction of stars  by the most recent rotation curve of the MW, while a link between mass to light ratio of nearby dSphs appears to be a complete mystery;
 \item explains the discrepant values of $M_{\rm tot}/L_{\rm V}$ of some dSphs because of the effect of distance that adequately reduces tidal shocks (e.g., Crater), of the impact of the Magellanic System (Carina II, Reticulum II, and Hydrus), and of the fact that the few anomalous dSphs (Sagittarius, Bootes, Crater II) are tidally stripped dSphs having passed well beyond their first pericenter;
 \item solves the question of disentangling dSphs from globular clusters, which share several common properties in stellar mass, as well as having a "connection" in the mass-radius plane; DM-free dSphs and globular clusters are both stellar systems dominated by tides of their host galaxy (e.g., Crater).
 \end{itemize}
 
In short, as far as we explore the consequences of the tidal shock scenario, it seems naturally nested into the physical properties of dSphs and the MW. The amplitude of the tidal shocks is consistent with the most accurate MW rotation curve, and N-body simulations show that it is caused by stars in resonance with the MW gravitational field variations when dSphs are near pericenters. An improved situation for their orbital motions is expected, in particular, having more precise tangential velocities from Gaia DR3 will significantly improve expectations from Figures~\ref{fig:1} and ~\ref{fig:2}. We can already make a few predictions from the present analysis, e.g., dSphs with small radii of about 20 pc that would lie farther than 100 kpc should have small velocity dispersions and then small DM mass-to-light ratios. We also suspect that the velocity dispersion measurement of Bootes II ($\sigma_{\rm los}$ = 10.5 $km s^{\rm -1}$ from five stars; see \citealt{Koch2009}) is  overestimated by a factor slightly larger than 2, which is in agreement with the discussion in \citet{Ji2016}. In the near future we would be able to predict velocity dispersions for all systems lying near their pericenter, which could be further tested through their consistency with the scaling relations shown in H19.\\

There is, however, a major argument against the tidal shock scenario, which comes from the $\Lambda$CDM cosmological model. If most MW dSphs were DM-free, the mismatch between DM halo and galaxy mass functions at the lower end (e.g., the so-called "missing dwarf problem") would become more than a major problem. This implies perhaps that the MW and its cortege of dSphs are unusual in the cosmological context. The tidal shock scenario may also suffer from bringing a complete change of paradigm for the nature of MW dSphs.  It is not fully excluded that the arguments listed above can be addressed by the DM-dominated scenario, though a demonstration of this is urgently needed. In the meantime, the tidal shock scenario appears the most successful and should be taken into account in any discussion on  the MW dSph properties. 

\section*{Acknowledgements}
We are grateful to Marcel Pawlowki for his comments and to Elisabetta Caffau for participating to one of our meetings. We warmly thank the referee for the useful and detailed comments that have improved the content of the manuscript.
This work was granted access to the HPC resources of MesoPSL financed
by the Region Ile de France and the project Equip at Meso (reference
ANR-10-EQPX-29-01) of the Programme Investissements d Avenir supervised
by the Agence Nationale pour la Recherche.
%The Acknowledgements section is not numbered. Here you can thank helpful colleagues, acknowledge funding agencies, telescopes and facilities used etc. Try to keep it short.

%%%%%%%%%%%%%%%%%%%%%%%%%%%%%%%%%%%%%%%%%%%%%%%%%%



%\bibliographystyle{mnras}


%%%%%%%%%%%%%%%%%%%%%%%%%%%%%%%%%%%%%%%%%%%%%%%%%%

%%%%%%%%%%%%%%%%% APPENDICES %%%%%%%%%%%%%%%%%%%%%

%\appendix


%Table 1 (to be built)

%%%%%%%%%%%%%%%%%%%%%%%%%%%%%%%%%%%%%%%%%%%%%%%%%%



\clearpage

\end{document}